\documentclass[twocolumn,showpacs,preprintnumbers,amsmath,amssymb]{revtex4}
\usepackage{graphicx}% Include figure files
\usepackage{dcolumn}% Align table columns on decimal point
\usepackage{bm}% bold math
\begin{document}
\title{Phase separation in the bosonic Hubbard model with ring
  exchange} 
\author{V. Rousseau, G. G. Batrouni}
\affiliation{
Institut Non-Lin\'eaire de Nice, Universit\'e de Nice--Sophia
Antipolis, 1361 route des Lucioles, 06560 Valbonne, France}
\author{R. T. Scalettar}
\affiliation{Physics Department,
University of California,
Davis CA 95616, USA}

\begin{abstract}
We show that soft core bosons in two dimensions with a ring exchange
term exhibit a tendency for phase separation.  This observation
suggests that the thermodynamic stability of normal bose liquid phases
driven by ring exchange should be carefully examined.
\end{abstract} 
% 05.30.Jp = Boson systems
% 03.75.Hh = Static properties of condensates; thermodynamical, 
%            statistical and structural properties
% 67.40.Kh = Boson degeneracy and superfluidity of He4: Thermodynamic properties
% 75.10.Nr 	Spin-glass and other random models
% 71.10.Fd 	Lattice fermion models (Hubbard model, etc.)
% 71.30.+h 	Metal-insulator transitions and other electronic transitions
% 02.70.Uu 	Applications of Monte Carlo methods 
\maketitle

Interest in ring exchange interactions in quantum many-body systems
has a long history, both theoretically and
experimentally\cite{Thouless65}. Recently, the ring exchange
interaction has been invoked in an effort to understand various
aspects of high temperature superconductivity.  While the Heisenberg
model alone provides a rather accurate picture of magnetic excitations
in the parent compounds of the cuprate
superconductors\cite{Manousakis91}, estimates of the magnitude of the
ring exchange term are as high as one quarter of the exchange
coupling\cite{Coldea01,Katanin02,MullerHartmann02} and it therefore
has been of interest to understand how this term might modify magnetic
properties\cite{Roger89,Honda93,Lorenzana99,Coldea01,Matsuda00}.  Ring
exchange interactions have also been suggested as a likely candidate
to reconcile the properties of the underdoped pseudogap regime.  The
basic picture is that the ring exchange interaction can give rise to a
new normal ``Bose metal'' phase at zero temperature in which there are
no broken symmetries associated with superfluidity or charge density
wave phases, and in which the compressibility is also
finite\cite{Fisher02}.

With these motivations partly in mind, Sandvik {\it et
al}\cite{Sandvik02} studied the phase diagram of the two-dimensional
spin-$1/2$ $XY$ model with spin exchange interaction on a square
lattice,
\begin{equation} 
\label{hardham}
H = -J \sum_{\langle {\bf ij}\rangle} B_{{\bf ij}} -
K \sum_{\langle {\bf ijkl}\rangle}
P_{{\bf ijkl}}
\end{equation}
where
\begin{eqnarray}
\label{bij}
B_{{\bf ij}} &=& S_{\bf i}^+ S_{\bf j}^-+S_{\bf i}^-S_{\bf j}^+
=2(S_{\bf i}^xS_{\bf j}^x+S_{\bf i}^yS_{\bf j}^y),
\nonumber
\\
\label{pijkl}
P_{{\bf ijkl}} &=& S_{\bf i}^+ S_{\bf j}^-S_{\bf k}^+ S_{\bf l}^-
+S_{\bf i}^- S_{\bf j}^+S_{\bf k}^- S_{\bf l}^+
\end{eqnarray}
and $\langle {\bf ij} \rangle$ denotes nearest neighbours and $\langle
{\bf ijkl}\rangle$ are sites at the corners of a plaquette. As is well
known, for $K=0$ this model is exactly equivalent to the hard core
bosonic Hubbard model, at half filling, with no interactions apart
from the constraint on site occupations, and it has only a superfluid
phase. Sandvik {\it et al}\cite{Sandvik02} studied the phase diagram
as a function of $K$ using the Stochastic Series Expansion
algorithm\cite{Sandvik91}.  They found that as $K$ increases, the
superfluid density, $\rho_s$, decreases up to a critical value, $K_c$,
where a phase transition takes place, and $\rho_s$ goes to zero with
long range order appearing in the momentum $(\pi,0)$ and $(0,\pi)$
channels of the plaquette-plaquette correlation function.  This
indicates the existence of a phase in which plaquettes with large and
small values of ring exchange alternate with a striped pattern across
the lattice.  This phase is also an incompressible insulator.  For yet
larger $K$, charge density wave order is established in which the site
occupations are alternatingly large and small.

It is believed \cite{Fisher02} that when the hardcore constraint is
relaxed, the striped plaquette phase might evolve into a normal
compressible conducting ``Bose metal'' in which none of the order
parameters mentioned above is non-zero.  This suggestion leads us to
study here the phase diagram of the soft core bosonic Hubbard model at
half filling with ring exchange interaction,
\begin{eqnarray}
\label{softham}
H &=& -t\sum_{\langle {\bf i},{\bf j}\rangle}(a_{{\bf
i}}^{\dagger}a_{{\bf j}}+ a_{{\bf j}}^{\dagger}a_{{\bf i}})
+U\sum_{{\bf i}} n_{{\bf i}} (n_{{\bf i}}-1) \\ \nonumber
&&+K\sum_{\langle {\bf ijkl}\rangle} (a_{\bf i}^\dagger a_{\bf j}
a_{\bf k}^\dagger a_{\bf l} +a_{\bf i} a_{\bf j}^\dagger a_{\bf k}
a_{\bf l}^\dagger)
\end{eqnarray}
where the destruction and creation operators satisfy $[a_{\bf
i},a_{\bf j}^\dagger]=\delta_{{\bf ij}}$, $n_{\bf i}=a_{\bf i}^\dagger
a_{\bf i}$ is the number operator at site ${\bf i}$ and $U$ is the
onsite interaction strength. For our quantum Monte Carlo (QMC)
simulations we used the World Line algorithm with four-site
decoupling\cite{Loh85}.  We verified our code for $K=0$ by comparing
with existing results for hard and soft core bosons with and without
near and next near neighbor interactions. For the $K\neq 0$ case, we
compared with the hard core results of \cite{Sandvik02}.

Before discussing results for the full many-body system, it is
interesting to study the behaviour of two bosons, since the formation
of a bound state is closely related to the issue of phase separation.
Ring exchange, like an attractive potential, favors proximity of two
bosons, since the action of such a term is nonzero only when the two
bosons live on the same plaquette.  In Fig.~1 we show the average
separation $\langle \Phi_0 | r^2 |
\Phi_0 \rangle$ of two bosons, normalized to the number of sites $L^2$
on an $L$x$L$ lattice.  As $K$ increases, there is a crossover at $K/t
\approx 3$ from a regime where the boson separation grows linearly
with system size, so that the normalized separation is size
independent, to one in which the boson separation does not grow with
system size.

While we do not show the associated data, plots of $\langle \Phi_0 |
r^2 | \Phi_0 \rangle$ for different soft-core repulsion $U$ reveal
% that the average separation is quite insensitive to the value of $U$.
that the average separation is insensitive to the value of $U$.
This will obviously be true in the unbound regime at small $K$, since
the density is so dilute.  It is less clear that this should be so in
the bound regime at large $K$.  However, as can be seen from the data
in Fig.~1, the radius of the bound state is several lattice spacings
($\langle r^2 \rangle \propto 0.1 L^2$ whence $r \propto 0.3 L$), so
here too the effect of $U$ is expected to be relatively small.

\begin{figure}
\label{boundstate}
\includegraphics[width=2.5in,height=3.0in,angle=270]{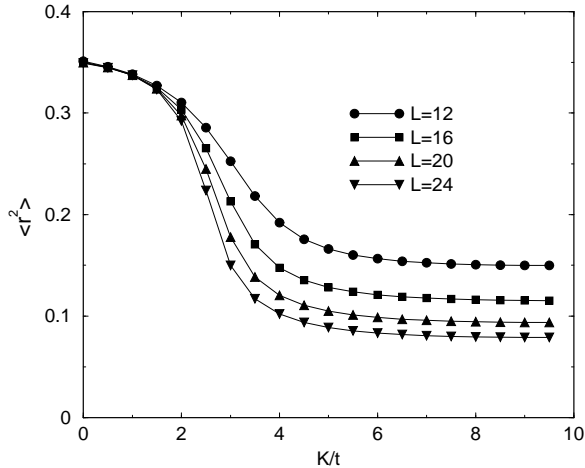}
\caption{ Average ground state separation $\langle \Phi_0 | r^2 |
\Phi_0 \rangle$ of two bosons, normalized to the system size, as a
function of the magnitude of the ring exchange energy scale $K$.  Here
the hopping $t=1$ and the soft-core repulsion $U=12$.  At small $K$,
the normalized separation is independent of lattice size, indicating
that the two bosons are spread independently 
throughout the lattice.  At larger $K$
the bosons prefer smaller separation to optimize the ring exchange
energy, indicating the formation of a bound state.}
\end{figure}

Fig.~1 suggests that there might be a tendency for ring exchange to
cause the bosons to clump together, and, in an extreme scenario, to
undergo phase separation.  However, at densities higher than the
dilute two boson case, this effect is opposed by the repulsion $U$.
The focus of this paper is to examine this competition and determine
the phase diagram of the soft core case as a function of both $U$ and
$K$ at half filling.

The most straightforward indication of phase separation comes from a
real space image of the boson density during the course of a
simulation.  Fig.~2 shows the average density distribution\cite{note2}
for $L=16$, $U=4$ and $K=2.5$.  We see indications that the bosons
undergo phase separation: At less than quarter filling (top panel) the
bosons clump together into a compact region of high density.  At
densities above quarter filling, on a lattice with periodic boundary
conditions, the number of occupied plaquettes is largest (and hence
the ring exchange energy is most negative) for a configuration where
the bosons stretch out in stripe across the lattice (bottom panel).

\begin{figure}
\label{phasesep1}
\includegraphics[width=3.0in,height=2.0in,angle=0]{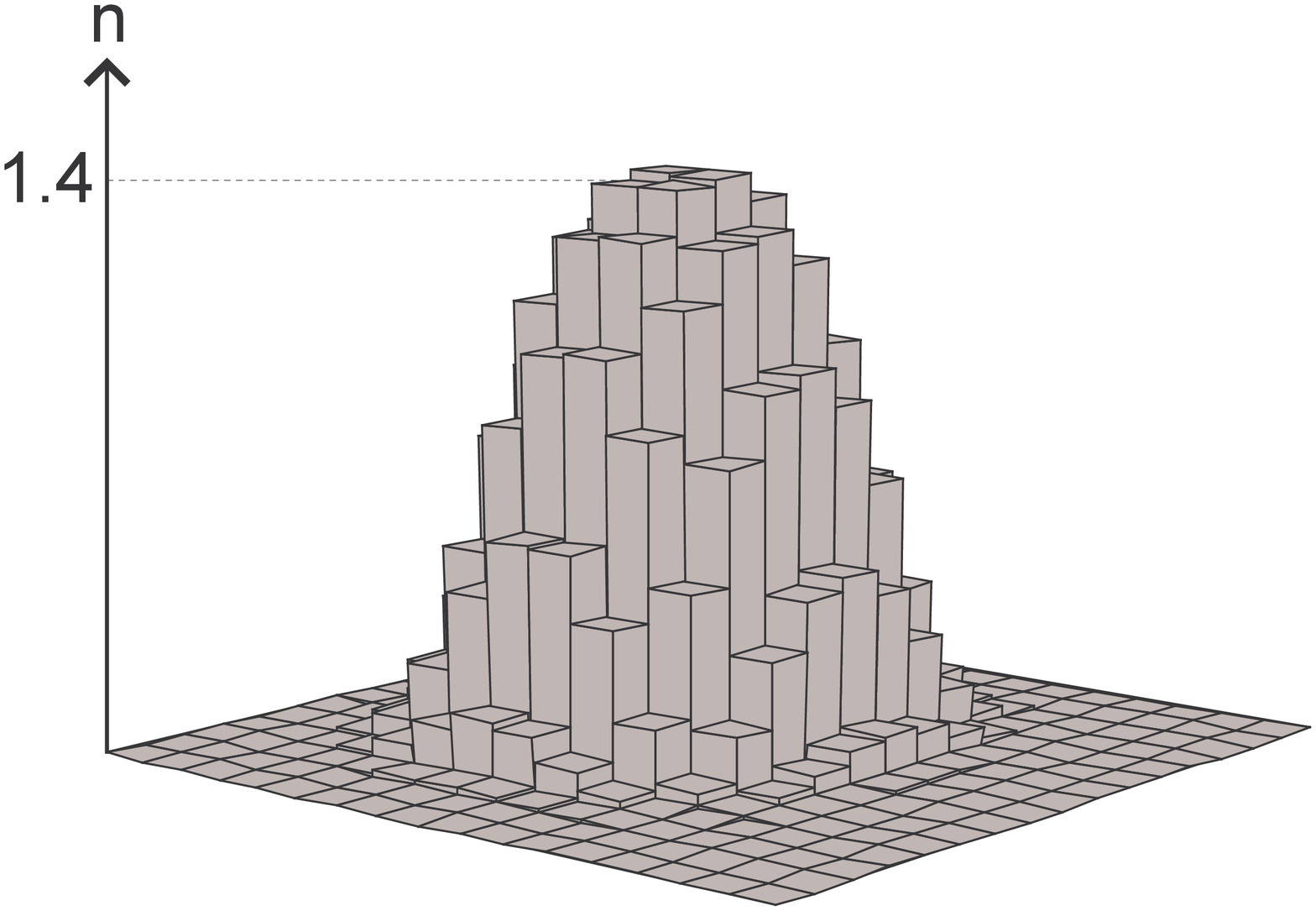}
\includegraphics[width=3.0in,height=2.0in,angle=0]{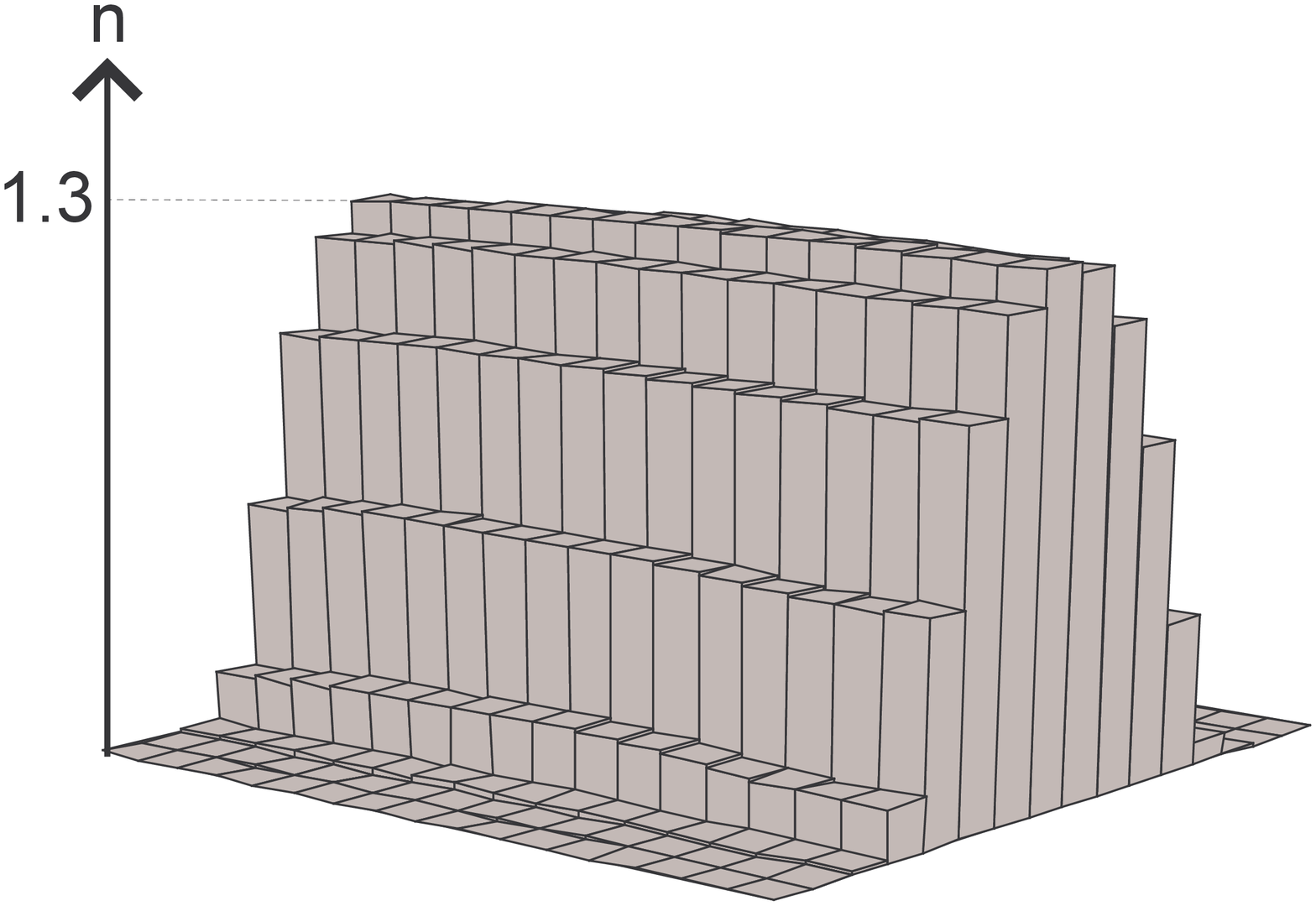}
\caption{Typical QMC results for the average density distribution in
the phase separated region.  Here $U=4$ and $K=2.5$.  Top panel:
$\rho$=50/256; Bottom panel: $\rho$=128/256 (half-filling).  At
half-filling, a stripe across our periodic boundary condition lattice
maximizes the number of occupied plaquettes, and hence minimizes the
ring exchange energy.}
\end{figure}

We will now demonstrate that phase separation is characteristic of a
large portion of the $K-U$ phase diagram by examining the
density-density correlation function and its associated structure
factor, fixing $U$ (or $K$) and scanning $K$ (or $U$).  As Fig.~2
illustrates, if the bosons phase separate, they may form a structure
in which a set of contiguous sites of about half the system size will
have appreciable boson occupation.  The other half of the lattice is
essentially empty. Therefore, if one examines the structure factor of
the density-density correlation function,
\begin{equation}
\label{structdenden}
S(k_x,k_y)\equiv {1 \over {L^2}} \sum_{{\bf r}} C({\bf r}){\rm e}^{-i{\bf
    r}.{\bf k}}
\end{equation}  
with 
\begin{equation}
\label{denden}
C({\bf r}) = {1\over {L^2}} \sum_{{\bf r}^{\prime}}\langle  n({\bf r}^{\prime})
n({\bf r}+{\bf r}^{\prime}) \rangle, 
\end{equation} 
one should observe a peak in $S$ at small momentum, e.g.~$(2\pi/L,0)$,
$(0,2\pi/L)$ or $(2\pi/L,2\pi/L)$ depending on the precise orientation
of the clump\cite{note3}.  By looking at the sum of the density
structure factor at these three smallest momentum values we are
sensitive to phase separation regardless of whether it occurs in a
puddle of roughly circular shape (Fig.~2a) or in some more elongated
pattern (Fig.~2b).  Fig.~3 clearly shows this behaviour: For $K<2$ at
$U=4$, $S$ is very small at the relevant momenta.  For $K>2$, phase
separation sets in.  Data for $16\times 16$ and $24\times 24$ lattices
are shown and their agreement indicates that this phase separation is
not a finite lattice effect.  The critical value of $K$ grows roughly
linearly with $U$.

It is also interesting to understand the behaviour of the superfluid
density $\rho_s$.  One does not necessarily expect $\rho_s$ to vanish
when phase separation occurs.  In fact, as is well known, $\rho_s \neq
0$ for the soft-core boson Hubbard model at all fillings, including commensurate density,
if $U/t$ is sufficiently small.  Similarly, here it is possible that
$\rho_s$ can survive in the dense region of the phase separated
lattice\cite{Fisher89,Batrouni90}. Our simulations show that when
phase separation first occurs, the populated region forms a band that
spans the whole system. The bosons may then delocalize along that
band, maintaining an (anisotropic) superfluid density. We have found
that when the bosons form such a band, the plaquette-plaquette
structure factor is also anisotropic and has long range correlations
along the direction of the band. As $K$ is increased further, the
populated region of the lattice takes the form of an island. In such a
case, the system may not be considered a superfluid in that one cannot
establish superflow across the system. However, the bosons may still
be delocalized over the extent of the island\cite{note4}.  Fig.~4
shows $\rho_s$ versus $K$.  
% As $K$ is increased for a fixed $U$,
% $\rho_s$ starts to decrease after peaking at a well defined value of
% $K$.  The value of $K$ at which $\rho_s$ takes on its maximal value
% increases as $U$ is increased, {\it i.e.} as the system approaches the
% hard core limit $U=\infty$.

\begin{figure}
\label{structfact}
\includegraphics[width=3in,height=2.5in,angle=0]{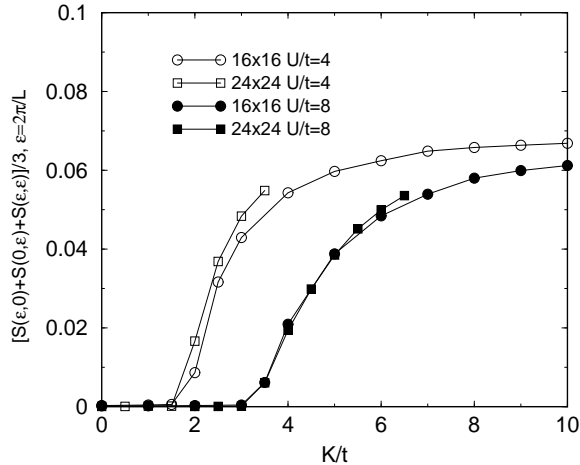}
\caption{The average structure factor, $(S(2\pi/L,0)+S(0,2\pi/L) +
S(2\pi/L,2\pi/L))/3$ versus $K$. We see a sharp increase in $S$ at $K=2$
for $U/t=4$ and $K\approx 4$ for $U/t=8$. The simulations were done
for $\beta=8$, and $\rho=0.5$. }
\end{figure}

\begin{figure}
\label{rhosvsK}
\includegraphics[width=2.5in,height=3in,angle=-90]{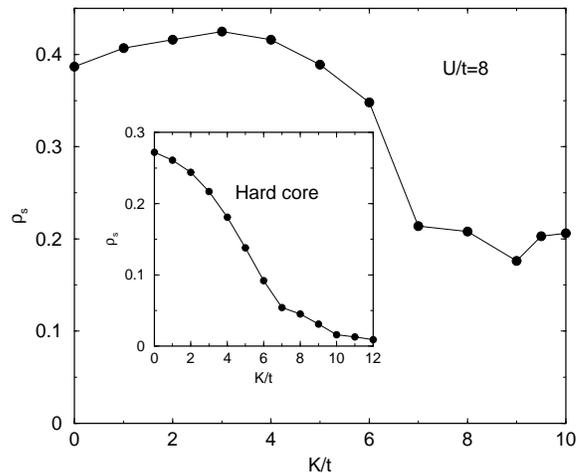}
\caption{
The superfluid density as a function of $K$ for $U=8$.  $\rho_s$
remains finite in the phase separated region, indicating that the
bosons are delocalized across the clump of occupied sites.  Inset: The
hard core limit for which, instead, $\rho_s \rightarrow 0$.  }
\end{figure}

By making scans like those of Fig.~3 at several values of $U$, we
construct the phase diagram which we show in Fig.~5.  Above the solid
line, the system undergoes phase separation.

\begin{figure}
\label{phasediag}
\includegraphics[width=3.0in,angle=0]{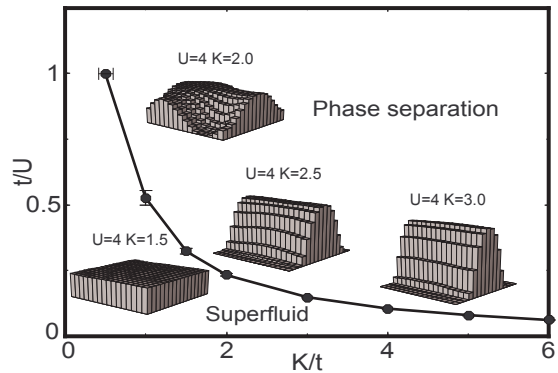}
\caption{ The phase diagram of the half-filled Bose Hubbard model
in the $U-K$ plane. Below the solid line, the system is superfluid
while above the line it phase separates. See text.}
\end{figure}

How do these results connect with the previous studies in the
hard-core limit?  There we know that a phase transition occurs at
$K_c/J \approx 7.9$ from a superfluid to a striped plaquette
phase\cite{Sandvik02}. We show in Fig.~6 the behaviour of the boson
density $\rho$ as a function of chemical potential $\mu$ in this
hard-core limit.  The jump in $\mu$ across half-filling $\rho=0.5$
shows that the plaquette ordered phase has a gap to the addition of
bosons (``charge excitations''), a result which is in agreement with
Sandvik {\it etal} \cite{Sandvik02}. The slope of the $\rho$ versus
$\mu$ is the compressibility $\kappa$. Consequently, if this curve
``bends backwards'', the system is thermodynamically unstable and
undergoes phase separation\cite{Batrouni00}.  While the data are not
conclusive, we do see hints of an instability.  For
$|\rho-0.5| > 0.008$, the slope is finite and corresponds to a normal
superfluid. For $| \rho -0.5| < 0.008$, the slope is either very
large, or perhaps negative, indicative of phase separation.

\begin{figure}
\label{rhovsmu}
\includegraphics[width=3.2in,height=2.3in,angle=0]{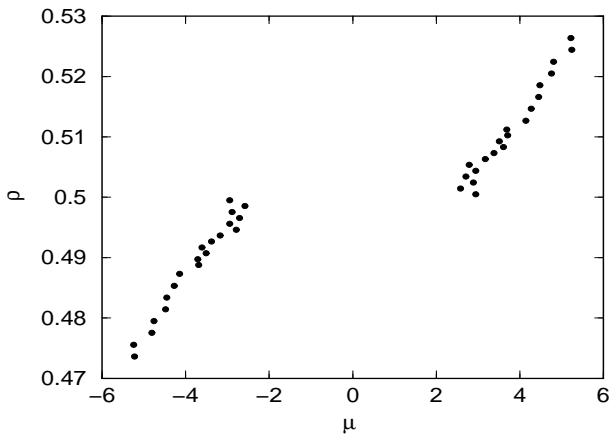}
\caption{ The boson density, $\rho$, versus the chemical potential,
$\mu$, for hard core bosons.  The jump in $\mu$ at $\rho=\frac12$ is
associated with the nonzero gap in the phase with long range striped
plaquette order.  There are indications of a region of negative
compressibility $\kappa = \partial \rho / \partial \mu$ immediately
adjacent to the gapped phase.  }
\end{figure}

So far we have addressed mostly the half filled case, and lower
densities.  It is of course of interest to examine higher fillings
where the effect of $U$ will be expected to discourage phase
separation. We have done simulations for $\rho =1$ and $\rho=1.5$ and
found in both cases that despite the higher densities, phase
separation still sets in above a critical value $K>K_{c}$ of the ring
exchange energy scale.  This result is not so surprising since the
on-site repulsion $U$ and the ring exchange term both scale with
density as $\rho^2$.  (Each of the four ring exchange
creation/destruction operators picks up a factor of $\approx \sqrt{n}$
when acting on a site with occupation $n$.)

In conclusion, we have shown that a sufficiently large ring exchange
energy can lead to a thermodynamic instability and phase separation.
We determined the critical $K$ as a function of the soft core
repulsion $U$ for a half-filled lattice and found roughly $K_c\approx
U/2$.  We conclude that the soft core boson Hubbard model does not
exhibit a normal Bose metal phase.  The bosons are either superfluid
or undergo phase separation. This phase separation also takes place
when the hopping parameter vanishes, $t=0$, a limit examined for the
quantum phase model in reference\cite{Fisher02} but which did not find
phase separation.

Finally, let us comment on the implications of our work for the
phase diagram of the spin-1/2 quantum 
Heisenberg model with a ring exchange term.
The kinetic energy term in the hard-core boson Hubbard model maps onto 
$ J \sum (S_{{\bf i}}^{x} S_{{\bf j}}^{x}
+ S_{{\bf i}}^{y} S_{{\bf j}}^{y})$
with exchange constant $J=2t$.
At the value $U=4t$ in our soft core model, double occupancy is
already very rare at half-filling, and hence we are almost in the
hard-core limit.
The value of the ring exchange energy scale required to drive phase
separation for this $U$ is $K \approx 2t$, or  
in other words, $K \approx J$.
To replicate the near-neighbor coupling of the $z$ components
of spin in the Heisenberg model 
we must include a near-neighbor repulsion in the bose-Hubbard model,
a term which clearly would suppress phase separation.
Thus we expect ring exchange to have the potential
to drive phase separation in the 
Heisenberg model only for $K$ considerably greater than $J$.

% The value of the ring exchange energy scale required to drive phase
% separation is $K \approx 2t$ at $U=4t$.  In spin language $K \approx
% J$ where $J$ is the coefficient of the $XY$ coupling, and thus the
% phase separation we observe is unlikely to be in the relevant
% parameter range of the cuprate superconductors.  On the other hand, it
% seems clear that this phase separation is germane to the interplay of
% ring exchange and the formation of a normal Bose liquid.  It will be
% interesting to address the existence of phase separation beginning
% from a boson hubbard model with a near-neighbor repulsion e.g. $V=2t$.
% At large $U$ this would correspond to a $J_z$ term in a Heisenberg
% Hamiltonian.  Such a term is, of course, likely to suppress phase
% separation.

{\it Note added:} After the completion of this work a preprint by
Melko, Sandvik and Scalapino (cond-mat/0311080) appeared where the
phase diagram of the doped hardcore system with ring exchange was
determined. 

\vskip0.2in
\noindent
\underbar{Acknowledgments:}
We acknowledge useful conversations with T. C. Newman.  This work was
supported by NSF-CNRS cooperative grant \#12929, NSF-DMR-0312261 and
NSF-INT-0124863.


\begin{references}
\bibitem{note3} When $U$ is reasonably large and all sites are either
empty or singly occupied, the structure factor at $S(k=(0,0))$ is
just the density.
\bibitem{note1} Reference \cite{Sandvik02} found that the plaquette
stripe order extends at least up to $K/J\simeq 12$ and perhaps up to
$K/J=16$. See \cite{Sandvik02} for more detail.
\end{references}
\end{document}